\documentclass[a4paper]{article}

\usepackage{INTERSPEECH2021}

\title{XMUSPEECH System for VoxCeleb Speaker Recognition Challenge 2021}
\name{Jie Wang$^1$, Fuchuang Tong$^1$, Zhicong Chen$^1$, Lin Li$^1$, Qingyang Hong$^2$, Haodong Zhou$^1$}
\address{
  $^1$School of Electronic Science and Engineering, Xiamen University, China\\
  $^2$School of Informatics, Xiamen University, China}
\email{\{lilin,qyhong\}@xmu.edu.cn}

\begin{document}

\maketitle
\begin{abstract}
  This paper describes the XMUSPEECH speaker recognition and diarisation systems for the VoxCeleb Speaker Recognition Challenge 2021. For track 2, we evaluate two systems including ResNet34-SE and ECAPA-TDNN. For track 4, an important part of our system is VAD module which greatly improves the performance. Our best submission on the track 4 obtained on the evaluation set DER 5.54\% and JER 27.11\%, while the performance on the development set is DER 2.92\% and JER 20.84\%.
\end{abstract}

\section{Track 2 System Description}

In this challenge, we focused on the Track 4 task, however, we also validated two speaker recognition systems  for the fully supervised speaker verification (open, Track2). The first system  was a 34-layer ResNet \cite{he2016deep} with a Squeeze-and-Excitation block  \cite{hu2018squeeze} (ResNet34-SE), and the second was ECAPA-TDNN \cite{desplanques2020ecapa}. Systems based on these two networks or their variations have achieved excellent results on last year VoxSRC 20, so we were interested in how they would perform this year.

For the ResNet34-SE network, the training data was VoxCeleb 2 dev part, and we used a 2-fold noise augmentation and 4-fold speed perturbation augmentation. The augmentation  was based on the Kaldi recipe. And the network was implemented on ASV-Subtools \cite{tong2021asv}. We used the SGD optimizer with an initial learning rate set to 0.01 and iteratively trained 6 epochs on a 12 GB-memory GPU.

For the ECAPA-TDNN system, we used the pre-trained system on SpeechBrain \cite{ravanelli2021speechbrain} directly. It was trained on the development set parts of VoxCeleb1 and VoxCeleb2, and online data augmentation was used during training. 
At the back end, features are scored using cosine similarity after subtracting the mean and length normalization.

\begin{table}[htbp]
  \centering
  \caption{Evaluation results with different model structures}
  \label{tab:sv_tab}
  \setlength{\tabcolsep}{1mm}{
    \begin{tabular}{rlrrr}
    \hline
          & \multicolumn{1}{c}{Feature} & \multicolumn{1}{c}{System} & \multicolumn{2}{c}{VoxSRC 2021 test} \\
          &       &       & \multicolumn{1}{r}{minDCF} & \multicolumn{1}{r}{EER} \\
    \hline
    1     &  Fbank 81 & \multicolumn{1}{l}{ResNet34-SE } & 0.303 & 5.13 \\
    2     & MFCC 80 & \multicolumn{1}{l}{ECAPA-TDNN } & 0.304 & 5.41 \\
    \hline
    3     & Fusion 1, 2  &       & 0.259 & 4.29 \\
    \hline
    \end{tabular}}
  \label{tab:addlabel}%
\end{table}%
The results of these two systems and their score fusions on the VoxSRC 21 test set are shown in Table~\ref{tab:sv_tab}. The results show that this year's test set was even more difficult, to the extent that last year's optimal system did not show competitive results on that test set. Instead of further tuning on Track 2 for the next challenge, we focused more on Speaker diarisation Track4 task.

\section{Track 4 System Description}

Our system was tuned on the development set of VoxConverse and reached a diarisation error rate (DER) of 2.92\%. Compared with the baseline system \cite{chung2020spot}, the DER of our system on VoxConverse evaluation set is 5.54\% which is improved by 69\%, while jaccard error rate (JER) is 27.11\%.

\subsection{System overview}
The speaker diarisation system consists of the following modules
\begin{itemize}
\item Voice activity detection (VAD).
\item Speaker embedding extractor.
\item Initial clustering modules.
\item Resegmentation modules.
\item Overlap speech detection (OSD).
\end{itemize}

\subsection{Voice activity detection}

We design three kinds of voice activity detection, which play a key role in the whole system. We evaluated the following voice activity detection schemes:
\begin{enumerate}
\item An energy-based system. We tune energy threshold of frames on VoxConverse development set.
\item A bidirectional long short-term memory network (Bi-LSTM) \cite{bredin2020pyannote} based system with two layers which is trained to output frames decisions. The module was pre-trained on the labeled DIHARD III development set and tuned on VoxConverse development set with 10 epochs.
\item A wav2vec2.0 \cite{baevski2020wav2vec} based automatic speech recognition (ASR) system. The system was pre-trained on 53k hours unlabeled data. Phoneme classes corresponding to silence and garbage were considered silence for the purpose of VAD and the rest of the classes were considered speech.
\end{enumerate}
Based on the above systems, we evaluate the performance with pyannote.metrics scoring tool \cite{bredin2017pyannote}. The main evaluation indexes were accuracy, precision, recall and detection error rate (DetER). Table~\ref{tab:VAD_tab} presents the comparison of results for each VAD system.

\begin{table*}[!htb]
\centering
\caption{Speaker 2nd indicates the second most probable speaker.“Bi-LSTM*” means the module was fine tuned on VoxConverse development set with 50 epochs, while the rest was fine tuned with 5 epochs.}
\label{tab:VBx_tab}
\begin{tabular}{cccccccccc}
\hline
System & VAD & Emb. & Clus. & Reseg. & Spk. 2nd & \multicolumn{2}{c}{dev} & \multicolumn{2}{c}{eval} \\ \cline{7-10} 
 &  &  &  &  &  & DER(\%) & JER(\%) & DER(\%) & JER(\%) \\ \hline
1 & \multicolumn{5}{c}{Baseline \cite{chung2020spot}} & - & - & 17.99 & 38.72 \\
2 & ASR & ResNet101 & AHC & VBx & VBx & 3.91 & 19.65 & - & - \\
3 & ASR & ResNet101 & AHC & VBx & heuristic & 3.82 & 19.21 & - & - \\
4 & Bi-LSTM & ResNet101 & NME-SC & VBx & VBx & 4.27 & 24.32 & - & - \\
5 & Bi-LSTM & ResNet101 & NME-SC & VBx & heuristic & 4.25 & 24.21 & - & - \\
6 & Bi-LSTM & ResNet101 & AHC & VBx & - & 7.06 & 21.57 & 7.79 & 25.76 \\
7 & Bi-LSTM & ResNet101 & AHC & VBx & VBx & 2.96 & 20.91 & - & - \\
8 & Bi-LSTM & ResNet101 & AHC & VBx & heuristic & 2.92 & 20.84 & 5.54 & 27.11 \\
9 & Bi-LSTM & ResNet152 & AHC & VBx & heuristic & 2.81 & 18.17 & - & - \\
10 & Bi-LSTM* & ResNet152 & AHC & VBx & heuristic & 2.41 & 18.02 & 5.62 & 25.19 \\ \hline
11 & \multicolumn{5}{c}{fusion of system 2,5,7,8} & 2.69 & 19.32 & 5.55 & 26.64 \\ \hline
\end{tabular}
\end{table*}

\begin{table}[]
\caption{The main evaluation indexes were accuracy, precision, recall and detection error rate (DetER).}
\label{tab:VAD_tab}
\begin{tabular}{ccccc}
\hline
System  & DetER(\%) & Acc.(\%) & Prec.(\%) & Rec.(\%) \\ \hline
Energy  & 5.65  & 94.74    & 95.72     & 98.76  \\
ASR     & 3.43  & 96.81    & 97.70     & 98.90  \\
Bi-LSTM & 2.47  & 97.70    & 98.29     & 99.26  \\ \hline
\end{tabular}
\end{table}

Bi-LSTM based system has the best performance when we removed silence shorter than 0.501s. Compared with systems that can be fine-tuned on the development set, ASR system has no development dataset to fine-tune. Considering this factor, the ASR system has great room for improvement. In subsequent experiments, we use Bi-LSTM based system to generate VAD label.

\subsection{Speaker embeddings}

The speaker embedding extractor is used to convert acoustic features into fixed dimensional feature vectors. We found that the Resnet34-SE and ECAPA-TDNN  extractor had inferior results to RensNet101 on the VoxSRC 21 Track4 Dev data, so we use ResNet101 as speaker embedding extractor whose inputs are 64 dimension log Mel filter bank features with 25ms frames and 10ms frames shift. In our system, the audio is cut into 1.44s segments with a 0.24s window shift. The 16 kHz x-vector extractor is trained using data from VoxCeleb1, VoxCeleb2 and CN-CELEB. We also apply ResNet152 as speaker embedding extractor that was trained on VoxCeleb1, VoxCeleb2 data set.

\subsection{Initial clustering}

In this stage, we evaluate different clustering algorithms, including agglomerative hierarchical clustering (AHC) and normalized maximum eigengap spectral clustering (NME-SC) \cite{park2019auto}. After comparison, AHC can achieve better performance when combined with VBx. The 256 dimensional speaker embedding is reduced to 128 dimensions by linear discriminant analysis (LDA). Then, the cosine similarity matrix is used as the input of the AHC model, and the threshold is set as -0.015. 

\subsection{Bayesian HMM for xvector clustering}

The detailed introduction of VBx model can be found in \cite{landini2022bayesian}, which was used to improve the performance of initial clustering. The results of the initial clustering are used for VBx model initialization. Variational Bayes HMM at x-vector level BHMM is used to cluster x-vectors. The HMM states represent speakers, the transition between states represent the speaker turns and the state distributions are derived from as PLDA model pre-trained on labeled x-vectors. The hyper-parameters mentioned in VBx model were tuned on Voxconverse development set so that Fa=0.15, Fb=5.5, loopP=0.33. In addition, the output of VBx model contains the second speaker label which can be applied to overlap speech detection.

\subsection{Overlap speech detection}

Voxconverse development set has two or more speakers speaking simultaneously. Considering the error of speech overlap is taken into account, we apply Bi-LSTM model as OSD. According to the output of OSD, we mask the second speaker labels. Bi-LSTM was trained on DIHARD III development set, and tuned on the development set of VoxConverse. Bi-LSTM takes overlap detection as a dichotomous classification problem that labels speech frames and non-speech frames. We set two thresholds to improve the robustness of the model. When the output probability of the model is higher than onset, the label changes from 0 to 1, while when the output probability of the model is lower than offset, the label changes from 1 to 0. Table~\ref{tab:OSD_tab} presents the comparison of results for each OSD system that was trained on different datasets. Due to the small amount of audio in the development set, we fine tune it five epochs to prevent over fitting.

\begin{table}
\caption{OSD performance on the development set. Evaluation of pre-trained OSD modelsm, in terms of detection error (DetER\%), accuracy, precision and recall.}
\label{tab:OSD_tab}
\setlength{\tabcolsep}{0.5mm}{
\begin{tabular}{llllll}
\hline
Pre-train & Tune & DetER(\%) & Acc.(\%) & Prec.(\%) & Reca.(\%) \\ \hline
AMI & - & 248.53 & 91.41 & 70.50 & 28.53 \\
DIHARD III & - & 89.50 & 96.88 & 53.93 & 72.10 \\
AMI & VoxC. & 54.02 & 98.12 & 77.16 & 65.33 \\
DIHARD III & VoxC. & 49.96 & 98.26 & 77.76 & 70.07 \\ \hline
\end{tabular}}
\end{table}

We also apply a heuristic algorithm \cite{otterson2007efficient} that considers the two closest speakers in times to infer a second speaker label. Table~\ref{tab:VBx_tab} presents the comparison of results for VBx and the heuristic algorithm.

We observed that the Bi-LSTM tuned with 50 epochs had obvious over-fitting. When applying heuristic algorithm to get second speaker label, the performance of system had been slightly improved. It seems that our systems has high similarity, so that the performance of our fusion system has not been improved. The fusion algorithm named DOVER-lap was mentioned in \cite{raj2021dover}.

\section{Conclusions}

This paper described the XMUSPEECH speaker recognition and diarisation systems for the VoxCeleb Speaker Recognition Challenge 2021. Our diarisation system consists of VAD, embedding extractor, initial clustering module, resegmentation module, and OSD. Finally, giving the credit to the VAD system, our best submission on the challenge obtained on the evaluation set DER 5.54\% and JER 27.11\%, while the performance on the development set is DER 2.92\% and JER 20.84\%. 


\bibliographystyle{IEEEtran}

\begin{thebibliography}{10}
\providecommand{\url}[1]{#1}
\csname url@samestyle\endcsname
\providecommand{\newblock}{\relax}
\providecommand{\bibinfo}[2]{#2}
\providecommand{\BIBentrySTDinterwordspacing}{\spaceskip=0pt\relax}
\providecommand{\BIBentryALTinterwordstretchfactor}{4}
\providecommand{\BIBentryALTinterwordspacing}{\spaceskip=\fontdimen2\font plus
\BIBentryALTinterwordstretchfactor\fontdimen3\font minus
  \fontdimen4\font\relax}
\providecommand{\BIBforeignlanguage}[2]{{%
\expandafter\ifx\csname l@#1\endcsname\relax
\typeout{** WARNING: IEEEtran.bst: No hyphenation pattern has been}%
\typeout{** loaded for the language `#1'. Using the pattern for}%
\typeout{** the default language instead.}%
\else
\language=\csname l@#1\endcsname
\fi
#2}}
\providecommand{\BIBdecl}{\relax}
\BIBdecl

\bibitem{he2016deep}
K.~He, X.~Zhang, S.~Ren, and J.~Sun, ``Deep residual learning for image
  recognition,'' in \emph{Proceedings of the IEEE conference on computer vision
  and pattern recognition}, 2016, pp. 770--778.

\bibitem{hu2018squeeze}
J.~Hu, L.~Shen, and G.~Sun, ``Squeeze-and-excitation networks,'' in
  \emph{Proceedings of the IEEE conference on computer vision and pattern
  recognition}, 2018, pp. 7132--7141.

\bibitem{desplanques2020ecapa}
B.~Desplanques, J.~Thienpondt, and K.~Demuynck, ``Ecapa-tdnn: Emphasized
  channel attention, propagation and aggregation in tdnn based speaker
  verification,'' in \emph{Interspeech2020}, 2020, pp. 1--5.

\bibitem{tong2021asv}
F.~Tong, M.~Zhao, J.~Zhou, H.~Lu, Z.~Li, L.~Li, and Q.~Hong, ``Asv-subtools:
  Open source toolkit for automatic speaker verification,'' in \emph{ICASSP
  2021-2021 IEEE International Conference on Acoustics, Speech and Signal
  Processing (ICASSP)}.\hskip 1em plus 0.5em minus 0.4em\relax IEEE, 2021, pp.
  6184--6188.

\bibitem{ravanelli2021speechbrain}
M.~Ravanelli, T.~Parcollet, P.~Plantinga, A.~Rouhe, S.~Cornell, L.~Lugosch,
  C.~Subakan, N.~Dawalatabad, A.~Heba, J.~Zhong \emph{et~al.}, ``Speechbrain: A
  general-purpose speech toolkit,'' \emph{arXiv preprint arXiv:2106.04624},
  2021.

\bibitem{chung2020spot}
J.~S. Chung, J.~Huh, A.~Nagrani, T.~Afouras, and A.~Zisserman, ``Spot the
  conversation: speaker diarisation in the wild,'' \emph{arXiv preprint
  arXiv:2007.01216}, 2020.

\bibitem{bredin2020pyannote}
H.~Bredin, R.~Yin, J.~M. Coria, G.~Gelly, P.~Korshunov, M.~Lavechin, D.~Fustes,
  H.~Titeux, W.~Bouaziz, and M.-P. Gill, ``Pyannote. audio: neural building
  blocks for speaker diarization,'' in \emph{ICASSP 2020-2020 IEEE
  International Conference on Acoustics, Speech and Signal Processing
  (ICASSP)}.\hskip 1em plus 0.5em minus 0.4em\relax IEEE, 2020, pp. 7124--7128.

\bibitem{baevski2020wav2vec}
A.~Baevski, H.~Zhou, A.~Mohamed, and M.~Auli, ``wav2vec 2.0: A framework for
  self-supervised learning of speech representations,'' \emph{arXiv preprint
  arXiv:2006.11477}, 2020.

\bibitem{bredin2017pyannote}
H.~Bredin, ``pyannote. metrics: A toolkit for reproducible evaluation,
  diagnostic, and error analysis of speaker diarization systems.'' in
  \emph{INTERSPEECH}, 2017, pp. 3587--3591.

\bibitem{park2019auto}
T.~J. Park, K.~J. Han, M.~Kumar, and S.~Narayanan, ``Auto-tuning spectral
  clustering for speaker diarization using normalized maximum eigengap,''
  \emph{IEEE Signal Processing Letters}, vol.~27, pp. 381--385, 2019.

\bibitem{landini2022bayesian}
F.~Landini, J.~Profant, M.~Diez, and L.~Burget, ``Bayesian hmm clustering of
  x-vector sequences (vbx) in speaker diarization: theory, implementation and
  analysis on standard tasks,'' \emph{Computer Speech \& Language}, vol.~71, p.
  101254, 2022.

\bibitem{otterson2007efficient}
S.~Otterson and M.~Ostendorf, ``Efficient use of overlap information in speaker
  diarization,'' in \emph{2007 IEEE Workshop on Automatic Speech Recognition \&
  Understanding (ASRU)}.\hskip 1em plus 0.5em minus 0.4em\relax IEEE, 2007, pp.
  683--686.

\bibitem{raj2021dover}
D.~Raj, L.~P. Garcia-Perera, Z.~Huang, S.~Watanabe, D.~Povey, A.~Stolcke, and
  S.~Khudanpur, ``Dover-lap: A method for combining overlap-aware diarization
  outputs,'' in \emph{2021 IEEE Spoken Language Technology Workshop
  (SLT)}.\hskip 1em plus 0.5em minus 0.4em\relax IEEE, 2021, pp. 881--888.

\end{thebibliography}


\end{document}